\documentclass[12pt]{article}
\usepackage{natbib,setspace,lscape,longtable}
\usepackage{natbib,epsfig,graphicx}
\usepackage{mathrsfs,amsmath,amsthm,amssymb,color, verbatim}
\usepackage{multirow,authblk}

\usepackage{geometry}
\usepackage{float}
\usepackage{array}
\usepackage{lscape}
\usepackage{bm}

\usepackage{comment}
\usepackage{setspace}
\usepackage{amsfonts}
\usepackage{indentfirst}
\usepackage{booktabs}
\usepackage{caption}
\usepackage{graphicx, subfig}

\numberwithin{equation}{section}

\newtheorem{theorem}{Theorem}
\newtheorem{lemma}{Lemma}
\newtheorem{remark}{Remark}
\newtheorem{proposition}{Proposition}
\newtheorem{corollary}{Corollary}

\usepackage{amsmath}
{
	\theoremstyle{plain}
	
}

\def\1{\mathbf{1}}

\def\mR{\mathbb{R}}

\def \tbe_t{\tilde{\bm{\epsilon}}_t}

\def\beq{\begin{equation}}
	\def\eeq{\end{equation}}
\def\ben{\begin{equation*}}
	\def\een{\end{equation*}}
\def\bea{\begin{eqnarray}}
	\def\eea{\end{eqnarray}}

\def\bda{\begin{eqnarray*}}
	\def\eda{\end{eqnarray*}}

\def\bet{\begin{theorem}}
	\def\eet{\end{theorem}}
\def\bel{\begin{lemma}}
	\def\eel{\end{lemma}}
\def\bep{\begin{proposition}}
	\def\eep{\end{proposition}}
\def\bec{\begin{corollary}}
	\def\eec{\end{corollary}}
\def\bg{\begin{figure}[tbph]\begin{center}}
		\def\eg{\end{center}\end{figure}}

\def\bc{\begin{center}}
	\def\ec{\end{center}}

\numberwithin{equation}{section}

\begin{document}

%

\title{\bf An Eigengap Ratio Test for Determining the Number of Communities in Network Data}
\author[1]{Yujia Wu}
\author[2]{Jingfei Zhang} 
\author[1]{Wei Lan}
\author[3]{Chih-Ling Tsai}

\affil[1]{School of Statistics and Center of Statistical Research, Southwestern University of Finance and Economics, Chengdu, China}
\affil[2]{Goizueta Business School, Emory University, Atlanta, GA}
\affil[3]{Graduate School of Management, University of California, Davis, CA}
\renewcommand*{\Authands}{ and }  
\date{}  
\maketitle


%


\begin{abstract}
	
To characterize the community structure in network data, researchers have introduced various block-type models, including the stochastic block model, degree-corrected stochastic block model, mixed membership block model, degree-corrected mixed membership block model, and others.
A critical step in applying these models effectively is determining the number of communities in the network. However, to our knowledge, existing methods for estimating the number of network communities often require model estimations or are unable to simultaneously account for network sparsity and a divergent number of communities. In this paper, we propose an eigengap-ratio based test that address these challenges. The test is straightforward to compute, requires no parameter tuning, and can be applied to a wide range of block models without the need to estimate network distribution parameters. Furthermore, it is effective for both dense and sparse networks with a divergent number of communities. We show that the proposed test statistic converges to a function of the type-I Tracy-Widom distributions under the null hypothesis, and that the test is asymptotically powerful under alternatives. Simulation studies on both dense and sparse networks demonstrate the efficacy of the proposed method. Three real-world examples are presented to illustrate the usefulness of the proposed test. \\

\end{abstract}


\section{INTRODUCTION}

Network data analysis has been extensively explored across various disciplines; see for example,  \cite{ko2009statistical}, \cite{goldenberg2010survey},\cite{valente2010social} and \cite{katona2011network}. Notably, \cite{lorrain1971structural} and \cite{granovetter1973strength} identified that the generation of network links often follows a block (or community) structure. Since then, numerous researchers have focused on block models to characterize the block structure of network generation mechanisms and enhance the utilization of network data. The widely studied block models include the stochastic block model \citep{holland1983stochastic,wasserman1994social,snijders1997estimation,nowicki2001estimation,bickel2009nonparametric,rohe2011spectral,choi2012stochastic,jin2015fast,zhang2016minimax,zhang2020mixed}, the degree-corrected stochastic block model \citep{karrer2011stochastic,zhao2012consistency,qin2013regularized,zhang2023adjusted,hu2021using,jin2023optimal}, the mixed membership model \citep{airoldi2008mixed,white2016mixed,ma2021determining} and the degree-corrected mixed membership model \citep{jin2017estimating,zhang2020detecting,ke2022optimal,han2023universal,jin2024mixed}, among others.

In this paper, we consider an undirected network with no self-loops, and assume that its associated adjacency matrix $A=(A_{ij})\in\mR ^{n\times n}$ is symmetric
with diagonal elements   all equal to zero. In addition, we assume that the network is generated from a Bernoulli-distributed random graph model. That is,
\beq\label{Ber}
A_{ij}\sim Bernoulli(P_{ij}), \quad\text{for every}\quad i<j,
\eeq
where $P=(P_{ij})\in\mR ^{n\times n}$ is the symmetric edge probability matrix. 
That is, $P_{ij}$ is the probability of connection between nodes $i$ and $j$, 
$i\neq j$. In existing block models, the probability matrix $P$ typically has a rank of $K$, where $K$ is the number of communities. We next briefly describe the four common block models mentioned above.

{\center\textbf{(a) The stochastic block model (SBM).}} The SBM \citep{holland1983stochastic} assumes that the network has $K$ communities and that each node belongs to one specific community. Let $g=(g(1),\cdots,g(n))^\top$ denote the membership vector, where $g(i)\in\{1, \cdots, K\}$ represents the community to which node $i$ belongs, for $i\in\{1, \cdots, n\}$. The SBM assumes that, for every $i\le j$,
\beq\label{def-SBM}
P_{ij}=Q_{g(i)g(j)},
\eeq
where $Q=(Q_{kl})\in\mR^{K \times K}$ is the $K\times K$ symmetric community probability matrix. The diagonal elements of $P$, $P_{ii}$, is set to be $Q_{g(i)g(i)}$ for any $i=1,\cdots,n$.
Under \eqref{def-SBM}, one can show that matrix $P$ is of rank $K$.

{\center\textbf{(b) The degree-corrected stochastic block model (DCSBM).}} The SBM assumes that the edge connecting any two nodes depends only on the communities in which they are located, and nodes within the same community have equal expected degrees. To allow for greater node degree heterogeneity, \cite{karrer2011stochastic} proposed the degree-corrected stochastic block model (DCSBM)
by introducing a vector of nodal degree parameters $\omega=(\omega_1,\omega_2,\cdots,\omega_n)^\top$ that measures the degree variation of nodes. Accordingly, the DCSBM assumes
\beq\label{def-DCSBM}
P_{ij}=\omega_i\omega_jQ_{g(i)g(j)},
\eeq
for every $i\leq j$, where $g$ and $Q$ are  defined in \eqref{def-SBM}.
The diagonal elements of $P$, $P_{ii}$, is set to be $\omega_i^2Q_{g(i)g(i)}$ for any $i=1,\cdots,n$.
Identifying DCSBM requires constraints such as $\sum_{i=1}^n\omega_iI\{g(i)=u\}=n_u$ for each community $1\le u\le K$, where  $n_u$ is the number of nodes in community $u$.
Under \eqref{def-DCSBM}, one can show that matrix $P$ is of rank $K$ \citep{jin2023optimal}.

{\center\textbf{(c) The mixed membership model (MM).}}
Both SBM and DCSBM require that each node belongs to a single community and the $K$ communities are nonoverlapping. To address this limitation, \cite{airoldi2008mixed} developed the mixed membership stochastic block model (MM) for handling the case of overlapping communities. For each node $i$, the MM introduces a $K$-dimensional probability mass vector, $\pi_i=(\pi_{i,1},\cdots,\pi_{i,K})^{\top}\in\mR^{K\times 1}$,  where $\pi_{i,k}$ represents the probability of node $i$ belonging to community $k$ for $1\le k\le K$ and $\sum_k\pi_{i,k}=1$. Then, the MM assumes that
\beq\label{def-MM}
P_{ij}=\pi_i^\top Q \pi_j,
\eeq
where $Q$ is defined in \eqref{def-SBM}. Under \eqref{def-MM}, one can show that matrix $P$ is of rank $K$.

{\center\textbf{(d) The degree-corrected mixed membership model (DCMM).}} 
\cite{jin2017estimating} introduced the degree-corrected mixed membership model (DCMM), which is an extension of DCSBM that accommodates overlapping 
communities. Specifically, DCMM assumes that
\beq\label{def-DCMM}
P_{ij}=\omega_i\omega_j\pi_i^\top Q \pi_j,
\eeq
where $\omega_i$s, $\pi_i$s and $Q$ are  defined in \eqref{def-SBM}-\eqref{def-MM}.
The DCMM is the most general class of models, and it includes the SBM, DCSBM and MM as special cases. One can show that the edge probability matrix $P$ under DCMM is also of rank $K$ \citep{jin2023optimal}.

Given the number of communities $K$, many algorithmic solutions have been proposed to estimate the community label $g$ or $(\pi_1,\ldots,\pi_n)$ in the above block models.  They include the modularity-based methods \citep{newman2004finding,newman2006modularity,sengupta2018block}, likelihood-based methods \citep{bickel2009nonparametric,amini2013pseudo,choi2012stochastic,zhao2012consistency,wang2023fast}, variational methods \citep{daudin2008mixture,airoldi2008mixed}, spectral clustering methods  \citep{rohe2011spectral,jin2015fast,lei2015consistency,joseph2016impact,qin2013regularized,sarkar2015role,sengupta2015spectral}, among others.

In practice, however, the number of communities $K$ is often unknown {\it a priori}. In recent years, many methods have been proposed for estimating $K$.
These methods include the spectral methods \citep{le2015estimating}, stepwise selection methods \citep{zhang2023adjusted,jin2023optimal}, sequential testing methods \citep{lei2016goodness,hu2021using}, network cross-validation methods \citep{chen2018network} and likelihood-based methods \citep{daudin2008mixture,wang2017likelihood,saldana2017many,noroozi2019estimation,hu2020corrected,ma2021determining}.
\cite{han2023universal} proposed a sequential test that can estimate matrix ranks and it can be applied to infer $K$ in block models. To our knowledge, these methods encounter at least one of the following challenges or restrictions.

First, the networks need to be assumed dense. For example, \cite{lei2016goodness} and \cite{hu2021using} assumed that the edge probabilities $Q_{kl}$s have a constant order of $O(1)$ for every $1\le k,l \le K$.
Second, the number of network communities $K$ needs to be finite as $n$ tends to infinity; see, for example, \cite{zhang2023adjusted} and \cite{han2023universal}.
Third, the unknown network distribution parameters need to be estimated. For example, \cite{lei2016goodness} and \cite{hu2021using} estimated $Q$, $\omega$ and $g$ before applying their methods to estimate $K$.
Similar approaches can be found in the literature such as \cite{noroozi2019estimation, zhang2023adjusted, hu2021using,ma2021determining,jin2023optimal}. Fourth, tuning parameters are needed; see, for example, \cite{wang2017likelihood}, \cite{hu2020corrected}, \cite{ma2021determining} and Han et al. (2023). 

\makeatletter\def\@captype{table}\makeatother
\setlength{\abovecaptionskip}{1cm}
\setlength{\belowcaptionskip}{0.2cm}
\begin{table}[!t]
	\setlength{\belowcaptionskip}{0.2cm}
	\scriptsize
	\centering
	\caption{Comparisons among our proposed method, \cite{lei2016goodness}, \cite{hu2021using}, \cite{han2023universal} and \cite{jin2023optimal}. The symbol ````$\checkmark$"" means applicable, and ``$\times$" stands for not applicable. }
	\label{Comparison}
\begin{tabular}{c|c|c|c|c|c}
	\hline {Feature}
	&\multicolumn{1}{c|}{Proposed method}&  \multicolumn{1}{c|}{Lei (2016)} & \multicolumn{1}{c|}{Hu et al. (2021)}& \multicolumn{1}{l}{Han et al. (2023)} & \multicolumn{1}{l}{Jin et al. (2023)}\\
	\cline{1-6}
	\hline
	Dense networks with $Q_{kl}=O(1)$ &$\checkmark$ &$\checkmark$ &$\checkmark$ &$\checkmark$ &$\checkmark$ \\
	Sparse networks with $Q_{kl}=o(1)$  &$\checkmark$ &$\times$ &$\times$ &$\checkmark$ &$\checkmark$\\
	Diverging number of communities &$\checkmark$ &$\checkmark$ &$\checkmark$ &$\times$ &$\times$ \\
	Avoiding parameter estimations &$\checkmark$ &$\times$ &$\times$ &$\checkmark$ &$\times$\\
	No tuning parameters &$\checkmark$ &$\checkmark$ &$\checkmark$ &$\times$ &$\checkmark$\\
	\bottomrule
\end{tabular}
\end{table}

In this paper, we propose an eigengap-ratio based test for estimating the rank of $P$, which can in turn be used to determine the number of communities in block models such as SBM, DCSBM, MM and DCMM.
The proposed test is straightforward to implement and requires only the calculations of the eigenvalues of $A$.
It accommodates sparse networks, and allows for a diverging $K$ without the need for parameter tuning; see Table \ref{Comparison} for a comparison of our proposed method with some recent methods.  
The main idea behind the proposed test statistic is as follows. Let $K$ and $K_0$ denote the true and hypothetical ranks of $P$, respectively, where $K$ is unknown.
To test whether $K=K_0$, we construct the test statistic as a ratio of two eigengaps, that is, the difference between pairs of eigenvalues derived from the adjacency matrix $A$. A key advantage of this approach is that it only requires a partial eigen-decomposition of the adjacency matrix $A$. Despite the simple form of the test statistic, deriving its null distribution is challenging, as $A_{ij}$s are Bernoulli random variables with distinct probability parameters $P_{ij}$s, which may tend to zero as $n$ increases.
 We show that the asymptotic null distribution of the test statistic follows a function of the type-I Tracy-Widom distribution constructed from the Wigner matrix. Additionally, we demonstrate that the proposed test is asymptotically powerful, with the test statistic diverging at a rate no slower than $n^{2/3}$ under the alternative hypothesis.

It is important to emphasize that our proposed test is designed to estimate the rank of the probability matrix $P$ and does not assume a specific block model or require the estimation of model parameters. 
Consequently, our test statistic does not provide a goodness-of-fit assessment for any specific block model. 
In contrast, block model-specific test methods such as those studied in \cite{lei2016goodness}, \cite{hu2021using}, \cite{ma2021determining} and \cite{jin2023optimal}, use test statistics to assess the goodness-of-fit of a presumed block model given a hypothetical number of 
communities. Due to the nature of their design, these tests typically rely on consistent estimators of the block model parameters, such as group membership $g$ or mixing proportions $\pi_i$s.

\section{METHOD AND THEORETICAL PROPERTIES}

\subsection{Hypothesis Testing}

Consider a network of $n$ nodes with the symmetric adjacency matrix $A$ generated from (\ref{Ber}). The aim of this paper is to test the following hypothesis:
\beq\label{test-hypo}
H_{0}: K=K_0 \quad \text{v.s.}\quad {H_{1}: K_0 < K\leq K_{\max}},
\eeq
where $K$ and $K_0$ denote the true and the hypothetical ranks of $P$,  respectively, and $K_{\max}$ is a pre-specified maximum value of $K$. In this paper, we set $K_{\max}=\max\{K_0+4, n^{2/5}\}$; see discussions after Condition (C2).  
We consider the one-sided alternative, $K>K_0$, since a network with $K$ communities can also be represented as a network with $K_0>K$ communities by splitting one or more true communities. The one-sided alternative was also considered in \cite{lei2016goodness}, \cite{chen2018network}, \cite{wang2017likelihood} and \cite{hu2021using}.

\subsection{An Eigengap-Ratio Based Test}

Recall $A=(A_{ij})\in\mR ^{n\times n}$ is the adjacency matrix with no self-loops. We define
\beq\label{eq:A1}
\tilde A=A-E(A),
\eeq
where $E(A)$ is the expected value of $A$, and $\tilde A$ is the noise matrix with zero-mean entries and zero diagonals elements. 
It is straightforward to see that entries of $\tilde A$ have finite variances. Given $E(A)=P-\text{diag}(P)$, (\ref{eq:A1}) can then be
rewritten as
\beq\label{eq:ARP}
A=P+\tilde A-\text{diag}(P),
\eeq
where $\text{diag}(P)$ represents the diagonal matrix constructed from $(P_{11},\cdots,P_{nn})^\top$.

Under the null hypothesis of $H_0$: $K=K_0$, the rank of $P$ is $K$, which implies that $P$ has $K$ nonzero eigenvalues.
Let $\lambda_k(A)$ denote the $k$-th largest eigenvalue of matrix $A$.
We can show that the eigenvalues $\lambda_k(A)$ for $k\ge K+1$ are determined by an $n-K$ principal submatrix of $\tilde A$, denoted by $\tilde A_{G_I}$, whose eigenvalues follow the type-I Tracy-Widom distribution.
This result motivates the proposal of an eigengap-ratio test statistic, defined as:
\beq\label{test-T}
T=\frac{\lambda_{K_0+1}(A)-\lambda_{K_{\max}+1}(A)}{\lambda_{K_{\max}+1}(A)-\lambda_{K_{\max+2}}(A)}.
\eeq
Given that the true rank is $K$, if $K_0<K\leq K_{\max}$, we expect  $\lambda_{K_0+1}(A)-\lambda_{K_{\max}+1}(A)$ to be much larger than $\lambda_{K_{\max}+1}(A)-\lambda_{K_{\max}+2}(A)$. In contrast, when $K_0=K$, the ratio of these two eigengaps is not expected to be large. In Theorem 1 of Section 2.3, we show that the asymptotic distribution of $T$ under $K_0=K$ is a function of the type-I Tracy-Widom distribution. Specifically, we first carefully bound the difference 
between $\lambda_k(A)$ and $\lambda_{k-K}(\tilde A_{G_I})$ for $k\ge K+1$, and then demonstrate that the distribution of $n^{2/3}(\sigma_2\lambda_{k-K}(\tilde A_{G_I})-L)$ converges to the type-I Tracy-Widom distribution, where $\sigma_2$ and $L$ are scaling and location parameters that do not depend on $k$.

The construction of $T$ in \eqref{test-T} offers two key advantages.
First, both $\lambda_k(A)-\lambda_{k-K}(\tilde A_{G_I})$ and the asymptotic distribution of $n^{2/3}(\sigma_2\lambda_{k-K}(\tilde A_{G_I})-L)$ involve scaling and location parameters that are very difficult to characterize or estimate. However, these parameters do not depend on $k$ and can be directly cancelled out in the eigengap-ratio statistic, eliminating the need for their estimation. 
Second, by using the eigengap ratio rather than individual eigenvalues, we avoid the requirement that $\lambda_k(A)$ converge to $\lambda_{k-K}(\tilde A_{G_I})$ (up to a scaling parameter) faster than $n^{2/3}$ for $k\ge K+1$. This simplifies the theoretical analysis and relaxes the assumptions necessary to establish the asymptotic null distribution.

\begin{remark}
Eigengap-ratio statistics have been studied in factor analysis and signal analysis. For example, \cite{onatski2009formal} proposed an  eigengap-ratio test to detect the number of factors by testing the rank of the factor loading matrix, and \cite{ding2022tracy} developed a maximum eigengap-ratio test to examine the number of signals by testing the rank of the signal matrix in a signal-plus-noise model.
However, in these models, tthe number of factors and signals are usually assumed to be fixed, and both the loading matrix and signal matrix are often presumed to be dense. In contrast, our test allows for a sparse $P$ and permits $K$ to diverge with $n$.
\end{remark}
\begin{remark}
The complexity of finding all $n$ eigenvalues of $A$ is roughly $O(n^3)$, and this can be computationally costly. In contrast, our test statistic $T$, for any given $K_{\max}$, only requires calculating the first $K_{\max}+2$ eigenvalues of $A$. Hence, the complexity can be reduced to $O(n^2K_{\max})$ or even to $O(K_{\max})$ if the network is sparse.
\end{remark}

\subsection{ Asymptotic Null Distribution}

The Tracy-Widom law, introduced by \cite{tracy1994level,tracy1996orthogonal} to characterize the distributions of the eigenvalues of Wigner matrices, has been extensively studied in a series of papers \citep{lee2014necessary,lee2015edge,schnelli2022convergence}. Specifically, let $W$ be a symmetric Wigner matrix wtih iid Gaussian entires of mean zero and variance $n^{-1}$. Then, $n^{2/3}(\lambda_1(W)-2)$ converges to the Tracy-Widom distribution. Under the null hypothesis $H_0$, we employ the Tracy-Widom law to derive the asymptotic null distribution of $T$. In addition, we reparameterize $P$ as $P=\rho_nP_n$, where $\rho_n\rightarrow 0$ and $P_n$ is fixed. This allows the connecting probability matrix to scale with $n$, a commonly approach in the literature \citep{bickel2009nonparametric,zhao2012consistency}.
We next present some regularity conditions.

\begin{itemize}
\item [(C1)] Assume the connecting probability matrix $P=(P_{ij})_{n\times n}$ is symmetric with $P_{ij}\in(0,1)$ for any $1\le i,j\le n$.
\item [(C2)] Assume $n^{2/3}\rho_n\to\infty$, and $K=O(n^{1/12-\psi})$, where $\psi$ is a constant satisfying $0<\psi\le 1/12$.
\item [(C3)]  Assume that $\tilde\lambda_{K}(P)\ge c^*_3n/K$, where $\tilde\lambda_K(P)$ is the $K$-th largest eigenvalue
of $P$ in absolute value and $c^*_3>0$ is a constant.
\end{itemize}

Condition (C1) requires the entries of the edge probability matrix $P$ to be in $(0, 1)$, a condition similarly considered in \cite{lei2016goodness}, \cite{wang2017likelihood} and \cite{hu2021using}.
Condition (C2) allows the edge probabilities $P_{ij}$s to decay to zero slower than $O(n^{-2/3})$, accommodating sparse networks. This condition is less restrictive than those of \cite{lei2016goodness} and \cite{hu2021using}, which assume dense networks where $P_{ij}=O(1)$. However, it is stronger than the conditions in \cite{han2023universal} and \cite{jin2023optimal} to ensure that the distribution of $n^{2/3}(\sigma_2\lambda_{k-K}(\tilde A_{G_I})-L)$ converges to the type-I Tracy-Widom distribution. It also allows the number of communities $K$ to diverge at a rate no faster than $O(n^{1/12})$, which is less stringent than the fixed $K$ considered in \cite{han2023universal}. 
By setting  $K_{\max}=\max\{K_0+4, n^{2/5}\}$, it holds that $K\le K_{\max}$. Condition (C3) imposes a lower bound on $\tilde\lambda_{K}(P)$, so it is bounded away from zero as $n\to\infty$. A similar condition was also considered in Theorem 3.1 of \cite{lei2015consistency}.

Based on the above conditions, we show the asymptotic null distribution of the test statistic $T$ in the following theorem.

\bet\label{theory-null}
Assume that Conditions (C1)-(C3) hold. Under the null hypothesis $H_0: K=K_0$, we have
$F_T(x){\sim} F_{T_W}(x)$ for any $x\in\mR$, as $n\to\infty$,
where $F_T(x)$ is the cumulative distribution function for the test statistic $T$, and $F_{T_W}(x)$ is the cumulative distribution function of $T_W$, defined as
$$
T_W=\frac{\lambda_1(W)-\lambda_{K_{max}-K_0+1}(W)}{\lambda_{K_{max}-K_0+1}(W)-\lambda_{K_{max}-K_0+2}(W)}.
$$
Here, $W$ is a Wigner matrix with i.i.d. entries of mean zero and variance $1/n$, and $\sim$ denotes asymptotic equivalence.
\eet

The above theorem indicates that the distribution of the test statistic $T$ is asymptotically equivalent to that of $T_W$. This allows us to employ a function of the Tracy-Widom distributions to test the null hypothesis $H_0$.
Specifically, for a given nominal level $\alpha$, let $c_\alpha=F_{T_W}^{-1}(1-\alpha)$ be the upper $\alpha$-quantile of $T_W$. We then tabulate the critical value $c_\alpha$ by approximating $F_{T_W}$ via a numerical method, such as Monte Carlo simulation, and reject  $H_0$ if $ T>c_\alpha$.

\subsection{Asymptotic Power}

Under the alternative hypothesis $H_1: K_0<K\leq  K_{\max}$, the eigenvalue $\lambda_{K_0+1}(A)$ is expected to be large, and the limiting distribution of $T$ no longer converges to that of $T_W$. Indeed, we demonstrate that $T$ diverges with $n$ under $H_1$, with strong discriminatory power against the alternative hypothesis.

\bet\label{theory-alter}
Under Conditions (C1)-(C3) and the alternative hypothesis $H_1: K_0<K\leq K_{\max}$, it holds that $P(T> Cn^{2/3})\to1$ as $n\to\infty$ for any positive constant $C$.
\eet

The above theorem implies that for a given nominal level $\alpha$, $P(T> c_\alpha)\to1$ under the alternative hypothesis $H_1$. In particular,
Theorem~\ref{theory-alter} indicates that $T$ diverges at a rate no slower than $n^{2/3}$ under the alternative, making our test more powerful than those of \cite{han2023universal} and \cite{hu2021using}. Specifically, the test statistic in \cite{han2023universal} is greater than a constant under the alternative, while the test statistic of \cite{hu2021using} diverges at an order of $O(\log n)$ under the alternative.
Additionally, the test statistic of \cite{hu2021using} lacks power under the planted partition stochastic block model, that is, a stochastic block model with equal community sizes, equal $Q_{ll}$s, and equal $Q_{lk}$s for $l\neq k$. We next demonstrate through simulation studies that $T$ performs well in both size and power.

\section{SIMULATION STUDIES}

To evaluate the performance of the proposed test  in finite samples,
we conduct simulation studies based on three models SBM, DCSBM and DCMM, to test the following hypotheses:
$$H_0: K=K_0\quad \text{v.s.}\quad H_1: K_0<K\leq K_{\max}.$$
The MM model is not included as it is a special case of DCMM.

To simulate from the DCMM in \eqref{def-DCMM} with $P_{ij}=\omega_i\omega_j\pi_i^\top Q \pi_j$,
we follow the approach in \cite{han2023universal} to generate $\pi_i$ and the approach in \cite{zhao2012consistency} to generate $\omega_i$.
Specifically,
let $\text{PM}_K=\{e_1,\cdots,e_K\}$,
where $K$ is the true number of communities and $e_k$ is an $n$-dimensional unit vector with the $k$th element equal to 1 and others equal to 0, and $\text{MM}_K=\Big\{(0.2,0.8,\underbrace{0,\cdots,0}_{K-2}),(0.8,0.2,\underbrace{0,\cdots,0}_{K-2}),(\underbrace{\frac{1}{K},\cdots,\frac{1}{K}}_K)\Big\}$. We generate $\omega_i$ independently from a distribution with unit expectation, that is,
$$
\omega_i=\left\{
\begin{array}{lr}
	\eta_i,& \text{w.p.}\quad 0.8,  \\
	9/11,& \text{w.p.}\quad 0.1, \\
	13/11,& \text{w.p.}\quad 0.1,
\end{array}
\right.
$$
where $\eta_i$ is uniformly distributed on the interval $[4/5,6/5]$. We further set $\Omega_1=I_n$ and $\Omega_2=\text{diag}\{\omega_1,\omega_2,\cdots,\omega_n\}$ to represent the two types of degree parameters, where $I_n$ is an $n$-dimensional identity matrix.

The network is set as $n=3,000$ with $K$ communities, the size of each community is $3000/K$, and $K\in\{3, 5, 10, 15, 20\}$. 
For the SBM setting, we set $\Omega=\Omega_1$ and $\pi_i\in \text{PM}_K$.
For the DCSBM setting, we set $\Omega=\Omega_2$ and $\pi_i\in \text{PM}_K$.
Under the DCMM setting, we set $\Omega=\Omega_2$ and $\pi_i\in \text{PM}_K\cup \text{MM}_K$.
This model assigns $M=n(K^{-1}-0.03)$ pure nodes to each community and then equally allocates $(n-MK)/3$ nodes  across the mixed membership with the three probability mass vectors in $\text{MM}_K$.

We consider both dense and sparse networks to assess the performance of the proposed test. We also compare the proposed test $T$ with five competing tests: two in  \cite{lei2016goodness}, one in  \cite{han2023universal} and two in  \cite{hu2021using}, denoted as  $T_{Lei}$,  $T_{Lei,Boot}$, $T_{Han}$, $T_{Hu}$, and $T^{aug}_{Hu,Boot}$, respectively. Here, $T_{Lei,Boot}$ and $T^{aug}_{Hu,Boot}$ are tests that apply a bootstrap correction and an augmented-bootstrap procedure  to $T_{Lei}$ and $T_{Hu}$, respectively. See more details in \cite{lei2016goodness} and \cite{hu2021using}. Since the methods of \cite{lei2016goodness} and \cite{hu2021using} are not applicable for DCMM, we omit them in simulation studies for DCMM. To obtain the limiting distribution $T_W$ in our test,  we generate 1000 Wigner matrices $W=(W_{ij})\in\mR^{n\times n}$, where the $W_{ij}$s are independently generated from $N(0,1/n)$ for $1\le i\le j\le n$. This allows us to approximate the distribution of $T_W$ and determine the critical value at a given nominal level.

\subsection{Simulations Under Dense Networks}

We consider two dense edge probability matrices, modified from \cite{hu2021using} and \cite{han2023universal}, denoted as $Q_1$ and $Q_2$, respectively. In $Q_1$, the edge probability between any two communities $u$ and $v$ is set to $0.1(1+4\times I(u=v))$. In $Q_2$, we define $Q_{2, kl}=0.1^{|k-l|}$ if $l\neq l$, and $Q_{2, kl}=(K+1-k)/K$ otherwise. The edge probabilities in $Q_1$ and $Q_2$ do not decrease with $n$, and represent dense network scenarios. We implement $Q_1$ for SBM and DCSBM, incorporating the methods of \cite{lei2016goodness}, \cite{hu2021using}, and \cite{han2023universal} for comparison. Additionally, we implement $Q_2$ for DCMM, and compare with \cite{han2023universal}.

Tables 2-3 report the empirical sizes and powers for SBM and DCSBM, respectively, under the edge probability matrix $Q_1$.
The simulation results reveal two important findings.
First, the empirical sizes of $T$, $T_{Han}$ and $T^{aug}_{Hu, boot}$ are close to the 5\% nominal level,
although $T^{aug}_{Hu, boot}$ performs slightly worse than $T$ and $T_{Han}$. Under SBM, when $K>5$, the sizes of $T_{Hu}$, $T_{Lei}$ and $T_{Lei, boot}$ deviate notably from the nominal level. Under DCSBM, when $K>5$, $T_{Hu}$ exhibits size distortions, which can be much larger than 5\%. Additionally, the empirical sizes of $T_{Lei}$ and $T_{Lei,boot}$ deviate significantly from the 5\% nominal level, except for $T_{Lei, boot}$ under $K=K_0=3$  in Table 3. This is because $T_{Lei}$ and $T_{Lei,boot}$ are not designed for DCSBM.
Second, the empirical powers of $T$ and $T^{aug}_{Hu, boot}$ are generally equal to 1, with $T$ performs more powerfully than $T^{aug}_{Hu, boot}$, while $T_{Hu}$ is less powerful. Notably, $T_{Han}$ shows almost no power.
This could be due to the fact that the assumptions in Theorem 3.2 of  \cite{han2023universal}  for ensuring the asymptotic power of $T_{Han}$ are not satisfied under the edge probability matrix $Q_1$. In addition, under SBM, the powers. of $T_{Lei}$, and $T_{Lei, boot}$ are generally equal to 1. However, under DCSBM, due to the size distortions of $T_{Lei}$ and $T_{Lei, boot}$, we do not further analyze their empirical powers.

Table 4 reports the empirical sizes and powers of DCMM under the edge probability matrix $Q_2$.
The simulation results show two findings. First, the empirical sizes of $T$ are close to the nominal level 5\%. However, $T_{Han}$ exhibits size distortion when $K>5$. This finding indicates that the test $T_{Han}$ may not be suitable for large $K$. 
Second, the empirical powers of $T$ get larger as $K-K_0$ increases. Since the sizes of $T_{Han}$ are distorted for $K>5$ and the powers of $T_{Han}$ are not steady for $K\leq 5$, we do not further analyze their empirical powers. In sum, our proposed test generally outperforms other competing methods under SBM, DCSBM and DCMM with dense networks.

\begin{table}[!t]
	\centering
	\caption{Dense SBM under $Q_1$. Proportion of rejections at nominal level $\alpha = 0.05$ for hypothesis test $H_0 : K = K_0$ v.s. $H_1 : K_0<K \leq K_{\max}$.}
\label{Tab:dense2-sbmQ1}
\renewcommand\arraystretch{1.5}{
	\scriptsize
	\begin{tabular}{c|ccccc|ccccc}
		\hline
		& \multicolumn{5}{c|}{$T$} &\multicolumn{5}{c}{$T_{Han}$} \\
		\cline{1-11}
		K &3 & 5& 10& 15 & 20 &3 & 5& 10& 15 & 20\\
		\hline
		$K_0=3$ &0.053&1&1&1&1    &0.052&0.068&0.048&0.035&0.025\\
		$K_0=5$ &*&0.060&1&1&1      &*&0.045&0.030&0.047&0.045\\
		$K_0=10$ &* &* &0.042&1&1          &* &* &0.047&0.030&0.043\\
		$K_0=15$ &*&* &*&0.058&1    &* &* &* &0.038&0.045\\
		$K_0=20$ &*&* &* &* &0.048    &*  &*  &*  &* &0.045 \\
		\hline
		&\multicolumn{5}{c|}{$T_{Hu}$}& \multicolumn{5}{c}{$T^{aug}_{Hu,boot}$}\\
		\cline{1-11}
		K &3 & 5& 10& 15 & 20 &3 & 5& 10& 15 & 20\\
		\hline
		$K_0=3$ &0.068&0.115&0.828&0.993&1  &0.067&0.993&1&1&1\\
		$K_0=5$ &*&0.080&0.580&0.976&1        &*&0.042&1&1 &1    \\
		$K_0=10$ &* &* &0.130&0.500&0.985      &* &* &0.067&1&1  \\
		$K_0=15$ &*  &*   &* &0.240&0.565    &*&* &* &0.035&0.915 \\
		$K_0=20$  &*    &* &*   &* &0.398      &*  &*  &*  &* &0.045 \\
		\hline
		&\multicolumn{5}{c|}{$T_{Lei}$}& \multicolumn{5}{c}{$T_{Lei,boot}$}\\
		\cline{1-11}
		K &3 & 5& 10& 15 & 20 &3 & 5& 10& 15 & 20\\
		\hline
		$K_0=3$ &0.065&1&1&1&1   &0.052&1&1&1&1 \\
		$K_0=5$ &*&0.072&1&1 &1          &*&0.097&1&1&1\\
		$K_0=10$ &* &* &0.182&1 &1     &* &* &0.137&1&1\\
		$K_0=15$ &*  &* &* &0.614&1        & *  &*  &* &0.731&1    \\
		$K_0=20$ &*    &* &*  &* &0.846     &*  &*   &*   &* &0.930  \\
		\bottomrule
\end{tabular}}
\end{table}

\begin{table}[!t]
\centering
\caption{Dense DCSBM under $Q_1$.  Proportion of rejections at nominal level $\alpha = 0.05$ for hypothesis test $H_0 : K = K_0$ v.s. $H_1 : K_0<K \leq K_{\max}$. }
\label{Tab:dense2-dcsbmQ1}
\renewcommand\arraystretch{1.5}{
	\scriptsize
	\begin{tabular}{c|ccccc|ccccc}
		\hline
		& \multicolumn{5}{c|}{$T$} &\multicolumn{5}{c}{$T_{Han}$} \\
		\cline{1-11}
		K &3 & 5& 10& 15 & 20 &3 & 5& 10& 15 & 20\\
		\hline
		$K_0=3$ &0.035&1&1&1 &1   &0.047&0.055 &0.044  &0.035&0.050\\
		$K_0=5$ &*&0.053&1&1&1      &*&0.065&0.050&0.055&0.044\\
		$K_0=10$ &*&*&0.048&1&1         &*&*  &0.053&0.050&0.050\\
		$K_0=15$ &*&*&*& 0.033 &1    &* &  *  &* & 0.050 &0.050 \\
		$K_0=20$ &*&*&*&* &0.043    &*   &*   &*   &* &0.052 \\
		\hline
		&\multicolumn{5}{c|}{$T_{Hu}$}& \multicolumn{5}{c}{$T^{aug}_{Hu,boot}$}\\
		\cline{1-11}
		K &3 & 5& 10& 15 & 20 &3 & 5& 10& 15 & 20\\
		\hline
		$K_0=3$ &0.035&0.275 &0.840&0.987&1  &0.028&0.855&1 &1&1\\
		$K_0=5$ &*&0.042&0.465&0.845&1          &*&0.053&0.892&1&1   \\
		$K_0=10$ &*&*  &0.183 &0.766&1    &*&* &0.071 &0.905&1  \\
		$K_0=15$ &*   &*    &* &0.302 &1     &*   &*   &* &0.062 &0.995 \\
		$K_0=20$  &*   &*   &*   &* &0.711    &*   &*   &*   &* &0.041\\
		\hline
		&\multicolumn{5}{c|}{$T_{Lei}$}& \multicolumn{5}{c}{$T_{Lei,boot}$}\\
		\cline{1-11}
		K &3 & 5& 10& 15 & 20 &3 & 5& 10& 15 & 20\\
		\hline
		$K_0=3$ &0.110&1&1 &0.995&0.990   &0.065&1&0.588 &0.573&0.645 \\
		$K_0=5$ &*&1&0.936&0.978&0.984   &*&0.007&0&0.008&0.688  \\
		$K_0=10$ &*&*  &0.267 &0.820&0.810   &*&*  &0.084 &0.012&0.006 \\
		$K_0=15$ &* &*   &* &0.115&0.612   &*   &*   &*   &0.015 &0   \\
		$K_0=20$ &*   &*   &*   &* &0.162   &*   &*   &*   &* &0.026   \\
		\bottomrule
\end{tabular}}
\end{table}

\begin{table}[!t]
\centering
\caption{Dense DCMM under $Q_2$.  Proportion of rejections at nominal level $\alpha = 0.05$ for hypothesis test $H_0 : K = K_0$ v.s. $H_1 : K_0<K \leq K_{\max}$. }
\label{Tab:dense2-dcmmQ2}
\renewcommand\arraystretch{1.5}{
	\scriptsize
	\begin{tabular}{c|ccccc|ccccc}
		\hline
		& \multicolumn{5}{c|}{$T$} &\multicolumn{5}{c}{$T_{Han}$} \\
		\cline{1-11}
		K &3 & 5& 10& 15 & 20 &3 & 5& 10& 15 & 20\\
		\hline
		$K_0=3$ &0.057&0.345& 0.610&0.865 &0.965  &0.033 &1&1&0.405&0.785\\
		$K_0=5$ &* &0.057 & 0.330& 0.440& 0.645& *&0.040& 1&0.405&0.785\\
		$K_0=10$ &* &* &0.037 &0.185 &0.395 &* & * &0.113&0.415&0.785\\
		$K_0=15$  & * & * & * &0.040 &0.185  &  *& * & * &0.343&0.780 \\
		$K_0=20$ & * & * & *& *&0.036  & * & * & *& *&0.680\\
		\bottomrule
\end{tabular}}
\end{table}

\subsection{Simulations Under Sparse Networks}

In this section, we consider two types of sparse edge probability matrix $\tilde Q_1$ and $\tilde Q_2$.
In $\tilde Q_1$, the edge probability between any two communities $u$ and $v$ is $n^{-5/9}(1+4\times I(u=v))$. In $\tilde Q_2$, we set $\tilde Q_{2, kl}=5n^{-5/9}0.1^{|k-l|}$ if $k\not = l$, and $\tilde Q_{2, kl}=5n^{-5/9}(K+1-k)/K$ otherwise.
Analogous to Section 3.1, we implement $\tilde Q_1$ for SBM and DCSBM and $\tilde Q_2$ for DCMM.

Tables 5-6 report the empirical sizes and powers for SBM and DCSBM, respectively, under the edge probability matrix $\tilde Q_1$.
The results reveal the following two important findings.
First, both $T$ and $T_{Han}$ control sizes well.
In addition, the sizes of $T_{Hu}$ and $T_{Lei}$ deviate notably from the nominal level.
This is likely due to the fact that the tests in \cite{lei2016goodness} and \cite{hu2021using} are not designed for sparse networks.
After applying a bootstrap correction and an augmented-bootstrap procedure,
although $T_{Lei,boot}$ and $T^{aug}_{Hu,boot}$  give several reasonable empirical sizes, though both tests lack theoretical justification in sparse networks.
Second, our proposed test $T$ is powerful against alternatives, and its empirical power steadily increases as $K-K_0$ gets large.
This is consistent with our theoretical findings. 
In contrast, the empirical powers of $T_{Han}$ are low since the two assumptions  in Theorem 3.2 of \cite{han2023universal} for ensuring the asymptotic power of $T_{Han}$ are not satisfied under $\tilde Q_1$. Since most of the empirical sizes of the tests proposed by \cite{lei2016goodness} and \cite{hu2021using} are distorted, we do not further analyze their empirical powers.

Under the sparse DCMM, the results in Table 7 show similar qualitative findings as in Table 4. For example, the empirical sizes of $T$ are close to the nominal level 5\% and the empirical sizes of  $T_{Han}$ are 
distorted when $K>5$. In addition, the empirical powers of $T$ get larger as $K-K_0$ increases. Since the sizes of $T_{Han}$ are distorted for $K>5$ and the powers of $T_{Han}$ are not steady for $K\leq 5$, we do not further analyze their empirical powers.

To make a more comprehensive comparison, we also carry out simulation studies with different network settings in the supplementary material. Specifically, we consider equal-sized communities with $n_k=300$ and the total number of nodes in the network is $n=Kn_k$, with $K\in\{2, 4, 6, 8, 10\}$. 
The simulation results of SBM and DCSBM under this network setting and the simulation results of DCMM under edge probabilities $Q_1$ and $\tilde Q_1$ are given in Tables S.1-S.6 of the supplementary material, which show similar findings as in Tables 2-7.
In summary, simulation studies demonstrate that our proposed test performs well across three block models under both dense and sparse networks. These findings are consistent with our theoretical results. 

\begin{table}[!t]
\centering
\caption{Sparse SBM under $\tilde Q_1$. Proportion of rejections at nominal level $\alpha = 0.05$ for hypothesis test $H_0 : K = K_0$ v.s. $H_1 : K_0<K \leq K_{\max}$.}
\label{Tab:sparse2-sbmQ1}
\renewcommand\arraystretch{1.5}{
	\scriptsize
	\begin{tabular}{c|ccccc|ccccc}
		\hline
		& \multicolumn{5}{c|}{$T$} &\multicolumn{5}{c}{$T_{Han}$} \\
		\cline{1-11}
		K &3 & 5& 10& 15 & 20 &3 & 5& 10& 15 & 20\\
		\hline
		$K_0=3$ &0.057&0.675&0.802&0.905&0.971    &0.052&0.058&0.034 &0.040  &0.050\\
		$K_0=5$ &*&0.047&0.772&0.820&0.956     &*&0.057&0.060&0.058&0.043\\
		$K_0=10$ &*&*&0.055& 0.540 &0.832         &*&*&0.038&0.035&0.038\\
		$K_0=15$ &* &*&* &0.053 &0.280    &*  &*  &* &0.047  &0.056\\
		$K_0=20$ &*&* &* &*  &0.033     &*  &*  &*  &* &0.040 \\
		\hline
		&\multicolumn{5}{c|}{$T_{Hu}$}& \multicolumn{5}{c}{$T^{aug}_{Hu,boot}$}\\
		\cline{1-11}
		K &3 & 5& 10& 15 & 20 &3 & 5& 10& 15 & 20\\
		\hline
		$K_0=3$ &0.247&0.858&0.992&0.990&0.995  &0.062&0.697&0.914&1&1\\
		$K_0=5$ &*&0.393&1&1&1                 &*&0.042&0.832&0.986&1 \\
		$K_0=10$ &*&*&0.990&1&1          &*&*&0.018&0.878 &1  \\
		$K_0=15$ &* &* &* &1 &1        &*&* &* &0.025 &0.996 \\
		$K_0=20$  &* &* &* &* &1      &* &* &* &* &0.085\\
		\hline
		&\multicolumn{5}{c|}{$T_{Lei}$}& \multicolumn{5}{c}{$T_{Lei,boot}$}\\
		\cline{1-11}
		K &3 & 5& 10& 15 & 20 &3 & 5& 10& 15 & 20\\
		\hline
		$K_0=3$  &0.993&1&1&1&1  &0.072&0.980&0.700&0.978&1 \\
		$K_0=5$   &*&1&1&1&1     &*&0.155&0.597&0.830&1 \\
		$K_0=10$  &*&*&1&1 &1      &*&*&0.010&0.924&1 \\
		$K_0=15$  &* &* &*  &1 &1     &* &* &* &0.072&1 \\
		$K_0=20$  &*  &*  &* &* &1     &*  &*  &* &*  &0  \\
		\bottomrule
\end{tabular}}
\end{table}

\begin{table}[!t]
\centering
\caption{Sparse DCSBM under $\tilde Q_1$.  Proportion of rejections at nominal level $\alpha = 0.05$ for hypothesis test $H_0 : K = K_0$ v.s. $H_1 : K_0<K \leq K_{\max}$. }
\label{Tab:sparse2-dcsbmQ1}
\renewcommand\arraystretch{1.5}{
	\scriptsize
	\begin{tabular}{c|ccccc|ccccc}
		\hline
		& \multicolumn{5}{c|}{$T$} &\multicolumn{5}{c}{$T_{Han}$} \\
		\cline{1-11}
		K &3 & 5& 10& 15 & 20 &3 & 5& 10& 15 & 20\\
		\hline
		$K_0=3$  &0.038&0.398&0.535&0.888&1     &0.042&0.050&0.048&0.033&0.043\\
		$K_0=5$  &*&0.057&0.243&0.492&0.898     &*&0.067&0.056&0.046&0.038 \\
		$K_0=10$  &*&*&0.046 &0.482&0.610        & *& *&0.047&0.036&0.040\\
		$K_0=15$  &* & *&*&0.040&0.530    &* &*& *&0.062&0.053\\
		$K_0=20$  &*&*&* &* &0.032     &* & *& *& *& 0.043 \\
		\hline
		&\multicolumn{5}{c|}{$T_{Hu}$}& \multicolumn{5}{c}{$T^{aug}_{Hu,boot}$}\\
		\cline{1-11}
		K &3 & 5& 10& 15 & 20 &3 & 5& 10& 15 & 20\\
		\hline
		$K_0=3$ &0.018&0.843&0.940&0.673&0.568  &0.060&0.475&0.786&0.873&1\\
		$K_0=5$ &*&0.138&1&0.932&1  &*&0.082&0.708&0.612&0.965 \\
		$K_0=10$ & *& *&0.850 &1&1      &* &*&0.053 &0.536&1 \\
		$K_0=15$ &* &* &*&1 &1        &* &* &*&0.151&0.910\\
		$K_0=20$  &* &* &* &*& 0.849   &* &* &* &* &0.016\\
		\hline
		&\multicolumn{5}{c|}{$T_{Lei}$}& \multicolumn{5}{c}{$T_{Lei,boot}$}\\
		\cline{1-11}
		K &3 & 5& 10& 15 & 20 &3 & 5& 10& 15 & 20\\
		\hline
		$K_0=3$  &0.173&1&1&1 &1  &0.040&0.963&0.560&0.745&0.350 \\
		$K_0=5$   &*&0.142&1&1&1  &*&0.157&0.491&0.161 &0.975 \\
		$K_0=10$  &* &*&0.826 &1&1     & * & *&0.077 &0.242&0.117 \\
		$K_0=15$  &* &* &*&0.036 &1    &* &* &*&0.036&1  \\
		$K_0=20$  &*  &*  &* &* &0     &*  &*  &* &*  &0  \\
		\bottomrule
\end{tabular}}
\end{table}

\begin{table}[!t]
\centering
\caption{Sparse DCMM under $\tilde Q_2$.  Proportion of rejections at nominal level $\alpha = 0.05$ for hypothesis test $H_0 : K = K_0$ v.s. $H_1 : K_0<K \leq K_{\max}$.}
\label{Tab:sparse2-dcmmQ2}
\renewcommand\arraystretch{1.5}{
	\scriptsize
	\begin{tabular}{c|ccccc|ccccc}
		\hline
		& \multicolumn{5}{c|}{$T$} &\multicolumn{5}{c}{$T_{Han}$} \\
		\cline{1-11}
		K &3 & 5& 10& 15 & 20 &3 & 5& 10& 15 & 20\\
		\hline
		$K_0=3$ &0.043&0.280 & 0.465& 0.655& 0.895 &0.047 &1&1&0.900& 0.960\\
		$K_0=5$ &* &0.037 & 0.250&0.455 &0.595 & *&0.040& 1&0.900 &0.960\\
		$K_0=10$ &* &* &0.050 &0.190 & 0.395&* & * &0.420&0.905&0.960\\
		$K_0=15$  & * & * & * &0.037 & 0.130 &  *& * & * &0.793&0.955 \\
		$K_0=20$ & * & * & *& *&0.041  & * & * & *& *&0.933\\
		\bottomrule
\end{tabular}}
\end{table}

\section{REAL DATA ANALYSES}

To illustrate the utility of our proposed method, we analyze three real-world networks. 
The first dataset is the facebook network in Simmons College that was first studied in \cite{traud2012social}. 
The Simmons College network contains 1,137 nodes, and its edge density is approximately 3.76\%. 
The second dataset is a political blog network that was first studied by \cite{adamic2005political} and later analyzed by \cite{lei2016goodness}, \cite{han2023universal} and \cite{hu2021using}. This network contains 1,222 nodes, and its edge density is approximately 2.24\%.  
The third dataset is a Sina Weibo user network that was analyzed by \cite{wu2022inward}.
This network has 2,580 nodes, and the edge density of this network is approximately 0.41\%.

\subsection{Simmons College network}
The Simmon college network contains all the friendship links between Facebook users within the Simmons College, recorded on a specific day in September 2005 \citep{traud2012social}. \citet{traud2012social} observed that the community structure of this network is strongly correlated with graduation year, meaning that students in the same year are more likely to be friends. Hence, the true number of clusters in this network can be considered as $K=4$. It was pointed out in \cite{jin2021improvements} that the community structure in the Simmons College network is very weak and community detection algorithms usally yield high misclassification errors.

We test $K_0=1,2,3,4,5,6$ using our proposed method. The resulting test statistics are $T=105.85, 58.49, 29.90, 24.19, 17.64$ and 14.25, with corresponding 5\% critical values $c_{0.05}=74.20, 82.99, 96.65, 87.62, 75.30$ and 63.99. Accordingly, $T$ rejects $H_0: K=1$ and does not reject $H_0: K=2$, $H_0: K=3$, $H_0: K=4$, $H_0: K=5$, and $H_0: K=6$. Hence, $T$ identifies at least two communities. We also employ $T_{Lei, boot}$, $T_{Hu,boot}^{aug}$ and $T_{Han}$ to conduct the same hypothesis testing. The test results are $T_{Han}=7.80, 1.62, -0.17, 1.28, 0.79$ and $-0.02$,  with the critical value $c_{0.05,Han}=1.65$. Therefore, $T_{Han}$ yields the same results as $T$, rejecting $H_0: K = 1$ but not $H_0: K = 2$, $H_0: K = 3$, $H_0: K = 4$, $H_0: K = 5$, and $H_0: K = 6$. Additionally, we find $T_{Lei, boot}=36.28, 23.81, 20.74, 6.56, 3.74, 2.68$, and $T_{Hu,boot}^{aug}>200$ for all $K_0=1,2,3,4,5, 6$, with correspondingly critical values $c_{0.05, Lei}=1.45$ and $c_{0.05,Hu}=3.41$, respectively. Therefore, $T_{Lei, boot}$ and $T_{Hu,boot}^{aug}$ reject all six null hypotheses.

\subsection{Political Blog network}

This dataset records links between internet blogs over the period of two months before the 2004 U.S. presidential election. The nodes are blogs, and the edges represent web links between the blogs. We consider the largest connected component of this network, consisting of 1,222 nodes, as commonly used in existing studies \citep{lei2016goodness,hu2021using,han2023universal}.
Note that all the blogs fall into two communities based on political stance, namely ``conservative" and ``liberal". 

Using this network dataset, \cite{lei2016goodness} tested for $H_0: K=2$ under the SBM.  In particular, he obtained $T_{Lei}=1172.30$ and $T_{Lei,Boot}=491.50$. Given the 5\% nominal level, both are much larger than the critical value $c_{0.05, Lei}=1.45$, and thus strongly reject the null hypothesis of $H_0: K=2$ under the SBM.
In \cite{hu2021using}, the test of $H_0: K=2$ was not rejected under the DCSBM. 
In addition, \cite{han2023universal}  tested $H_0: K=1$ and  $H_0: K=2$  separately. They found that $T_{Han}=2.71$ under $H_0: K=1$ and $T_{Han}=-0.89$ under $H_0: K=2$, where the critical value is $c_{0.05,Han}=1.65$. Accordingly, $T_{Han}$ rejected $H_0: K=1$ and did not reject $H_0: K=2$. Their result supports the finding in \cite{hu2021using}.

We also test $H_0: K=1$ and  $H_0: K=2$ using our proposed test. Under $H_0: K=1$, we have $T=34.25$, which is larger than the critical value $c_{0.05}=30.16$. Under $H_0:K=2$, we obtain that $T=6.84$, which is smaller than the critical value $c_{0.05}=22.18$. Consequently, we reject $H_0: K=1$ and do not reject $H_0: K=2$. This finding agrees with that of \cite{han2023universal}, which corroborates the prior knowledge that all blogs can be divided into politically ``conservative" and ``liberal" communities.

\subsection{Sina Weibo Data}
This dataset includes friendship links between 2,580 Sina Weibo users, and $A_{ij}=1$ if user $i$ and user $j$ follow each other \citep{wu2022inward}. According to the analysis of \cite{wu2022inward}, users can be partitioned into four quadrants based on two nodal influence indices. For each node $i$, the two indices are 'inward influence,' which measures node $i$'s receptivity to being influenced by others, and 'outward influence,' which gauges the influence node $i$ exerts on others.
We report the results of testing $K_0=1,2,3,4,5$, as all tests with $K_0>5$ yield the same conclusion. The resulting test statistics are $T=105.43, 42.83, 28.12, 24.25$ and 22.67, with corresponding 5\% critical values $c_{0.05}=76.22, 53.86, 50.11, 32.78$ and 32.04. Thus, $T$ rejects $H_0: K=1$ and does not reject $H_0: K=2$, $H_0: K=3$, $H_0: K=4$ and $H_0: K=5$. Hence, $T$ identifies at least two communities. We also employ $T_{Lei, boot}$, $T_{Han,boot}^{aug}$ and $T_{Han}$ to conduct hypothesis testing. The test results show $T_{Lei, boot}>100$ and $T_{Han,boot}^{aug}>200$ for all $K_0=1,2,3,4,5$ with their correspondingly critical values $c_{0.05, Lei}=1.45$ and $c_{0.05,Hu}=3.41$, respectively. For $T_{Han}$, the test statistics are 19.26, 7.85, 8.66, 8.92 and 3.65, with critical values $c_{0.05,Han}=1.65$. Therefore, $T_{Lei, boot}$, $T_{Han,boot}^{aug}$ and $T_{Han}$ reject all five null hypotheses.

\section{CONCLUDING REMARKS}

Two possible avenues for future research are identified. First, generalize the proposed
method to accommodate the nonparametric graphical models \citep{wolfe2013nonparametric}. Second, study the validity of  the  eigengap-ratio test for correlated binary data such as the network generated by a notable latent space model \citep{hoff2002latent}. We believe these efforts will further increase the applicability of the proposed method.

\section*{SUPPLEMENTARY MATERIAL}

The supplementary material includes five sections. Section S.1 provides technical notation. Section S.2 introduces a lemma used to prove Theorem~1. Sections S.3--S.4 present the proofs of Theorem~1 and Theorem~2, respectively. Section S.5 reports the additional simulation results. 



\bibliographystyle{Chicago}
\bibliography{paper-ref}

\end{document}